\def\wordcount{1}		

\ifdefined\preprint
  \documentclass[preprint,review,12pt]{elsarticle}
\fi
\ifdefined\wordcount
  \documentclass[final,3p,times,twocolumn]{elsarticle}
\fi
\ifdefined\final
  \documentclass[final,3p,times,twocolumn]{elsarticle}
\fi

\usepackage{graphicx,stfloats}
\usepackage{color}

\usepackage{graphicx}
\usepackage{longtable}
\usepackage{tabularx}
\usepackage{url}
\usepackage{subcaption}
\usepackage{lipsum} 
\usepackage{amssymb,amsmath}
\usepackage[version=3]{mhchem}

\newcommand{\ave}[1]{\langle \, #1 \, \rangle}

\newcommand{\Rel}{\mathrm{Re}_{\lambda}}
\newcommand{\Ret}{\mathrm{Re_T}}
\newcommand{\Ka}{\mathrm{Ka}}
\renewcommand{\Re}{\mathrm{Re}}

\definecolor{myred}{rgb}{0.8,0,0}
\definecolor{myblue}{rgb}{0,0,1}
\definecolor{mygreen}{rgb}{0,0.39,0}

\newcommand{\added}[1]{#1}

\newcommand{\addedfig}[1]{#1}

\newcommand{\removed}[1]{}

\def\one{67mm}
\def\two{144mm}

\biboptions{sort&compress}

\journal{Proceedings of the Combustion Institute}

\begin{document}

\begin{frontmatter}

\title{Turbulent flame speed and reaction layer thickening in premixed jet flames at constant Karlovitz and 
increasing Reynolds numbers}

\author[rwth]{Antonio Attili\corref{cor1}}
\ead{a.attili@itv.rwth-aachen.de}
\author[kaust]{Stefano Luca}
\author[rwth]{Dominik Denker}
\cortext[cor1]{Corresponding author}
\author[ut]{Fabrizio Bisetti}
\author[rwth]{Heinz Pitsch}

\address[rwth]{Institute for Combustion Technology, RWTH Aachen University, 52056 Aachen, Germany}
\address[kaust]{King Abdullah University of Science and Technology (KAUST),
                Clean Combustion Research Center (CCRC), 23955 Thuwal, Saudi Arabia}
\address[ut]{Department of Aerospace Engineering and Engineering Mechanics,
             University of Texas at Austin, Austin, TX 78712, USA}

%
%
%
%

\begin{abstract}
A series of Direct Numerical Simulations (DNS) of lean methane/air flames
was conducted to investigate the enhancement of the
turbulent flame speed and  modifications to the reaction layer structure 
associated with the systematic increase of the integral scale of turbulence $l$ while the Karlovitz number and the Kolmogorov scale
are kept constant.
Four turbulent slot jet flames are simulated at increasing Reynolds number and
up to $\Re \approx 22000$, defined with the bulk velocity, slot width, and the reactants' properties. 
The turbulent flame speed $S_T$ is evaluated locally at selected streamwise locations
and it is observed to increase both in the streamwise direction for each flame and across flames for
increasing Reynolds number, in line with a corresponding increase of the turbulent integral scale.
In particular, the turbulent flame speed $S_T$ increases exponentially with the integral scale 
for $l$ up to about 6 laminar flame thicknesses, while the scaling becomes a power-law
for larger values of $l$.  
These trends cannot be ascribed completely to the increase in the flame surface,
since the turbulent flame speed looses its proportionality to the flame area as the integral scale increases;
in particular, it is found that the ratio of turbulent flame speed to area attains a power-law scaling $l^{0.2}$. 
This is caused by an overall broadening of the reaction layer for increasing integral scale,
which is not associated with a corresponding decrease of the reaction rate, causing
a net enhancement of the overall burning rate. 
This observation is significant since it suggests that a continuous increase in the size of the largest scales of turbulence
might be responsible for progressively stronger modifications of the flame's inner layers
even if the smallest scales, i.e., the Karlovitz number, are kept constant.

\end{abstract}

\begin{keyword}

Turbulent premixed flames \sep Direct Numerical Simulation \sep Turbulent flame speed \sep High Reynolds number \sep Flame Thickening

\end{keyword}

\end{frontmatter}

\ifdefined \wordcount
\clearpage
\fi

\section{Introduction}

The propagation of premixed flames subject to the stirring and straining of a turbulent velocity
field is known to be affected by two fundamental mechanisms~\citep{peters2000turbulent,veynante2002turbulent,driscoll2008turbulent,driscoll2020premixed}.
Turbulence causes an increase of the area of the flame surface, contributing to
an enhancement of the overall burning rate. In addition, turbulence can change the flame structure by perturbing the balance between reaction and diffusion,
modifying the local burning rate, sometimes to the point of inducing local extinction, and 
variation of the diffusive fluxes, which are associated with perturbations to the local thickness of the 
different layers of the flame structure. 
A commonly accepted hypothesis is that the perturbation 
of the inner layer of the flame depends only on the Karlovitz number $\Ka$, which parametrizes the ratio between the
flame thickness $\delta_L$ and the size of the smallest turbulent scales $\eta$, while variations of Reynolds number $\Re$ and
turbulence integral scale $l$ do not play a role if the Karlovitz number is kept constant~\citep{peters2000turbulent}.

In this work, we analyze the turbulent flame speed and the modifications of the inner reaction layer in a series of four jet flames characterized by 
nearly constant Karlovitz number and turbulent intensity (velocity root mean square) $u'$,
while the Reynolds number and turbulence integral scale vary significantly across the
flame series.
In particular, a ratio of integral scale, computed in the unburned gas, to laminar flame thermal thickness 
of $\approx 20$ was achieved at the largest jet Reynolds number of 22400. 
A previous analysis of the same Direct Numerical Simulation (DNS) dataset~\citep{luca2018statistics}
focused on the growth rates of the area of the flame surface and demonstrated the scaling of flame stretch, and
its components, with the Kolmogorov scale. 

In addition to the relative scarcity of studies that isolated the effects of the 
integral scale on the turbulent flame speed and flame structure~\citep{peters1999turbulent,lapointe2016simulation,luca2018statistics,kulkarni2020reynolds}, 
this analysis is motivated by the recent effort of \citet{skiba2018premixed} and \citet{driscoll2020premixed}
who demonstrated a transition from a flamelet regime 
towards a behavior characterized by thicker preheat layers for integral scales of increasing size
at constant $u$', i.e., moving horizontally in the Borghi-Peters diagram of turbulent premixed combustion~\citep{peters2000turbulent}.
Starting from these observations, we investigate whether an increase in the size of the largest turbulent scale (a Reynolds number increase)
promotes a modification to the
structure of the flame, in terms of modification of the local burning rate and inner layer thickness,
for a fixed size of the smallest scale relative to the laminar flame thickness (constant Karlovitz number).
In addition, the spatial inhomogeneity and
streamwise evolution of the turbulent flame speed and flame thickening
is assessed.

\section{Configuration, numerical methods, and overview of the flames}\label{sec:conf}

A slot turbulent premixed jet flame with equivalence ratio $\phi=0.7$, temperature 800~K and pressure of 4 atm,
surrounded by a coflow of burnt gases, is considered.
A summary of all relevant flow parameters is provided in Tab.~\ref{table}.
The database is described in details in Ref.~\citep{luca2018statistics};
therefore, only a breif summary is provided here.
Based on one-dimensional simulations of a freely propagating flame, the
laminar flame speed is $S_L=1$~m~s$^{-1}$ and the thermal thickness is $\delta_L=110$~$\mu$m.
The bulk velocity of the jet is $U$ = 100~m~s$^{-1}$ and the coflow has a uniform velocity of 15~m~s$^{-1}$.
For the four cases considered,
the Reynolds number $\Re = UH/\nu$ varies from 2800 to 22400 as the slot's width $H$ increases
from 0.6 to 4.8 mm ($\nu$ is the kinematic viscosity of the reactants).

\begin{figure}[]
\addedfig
\centering \includegraphics[width=\one]{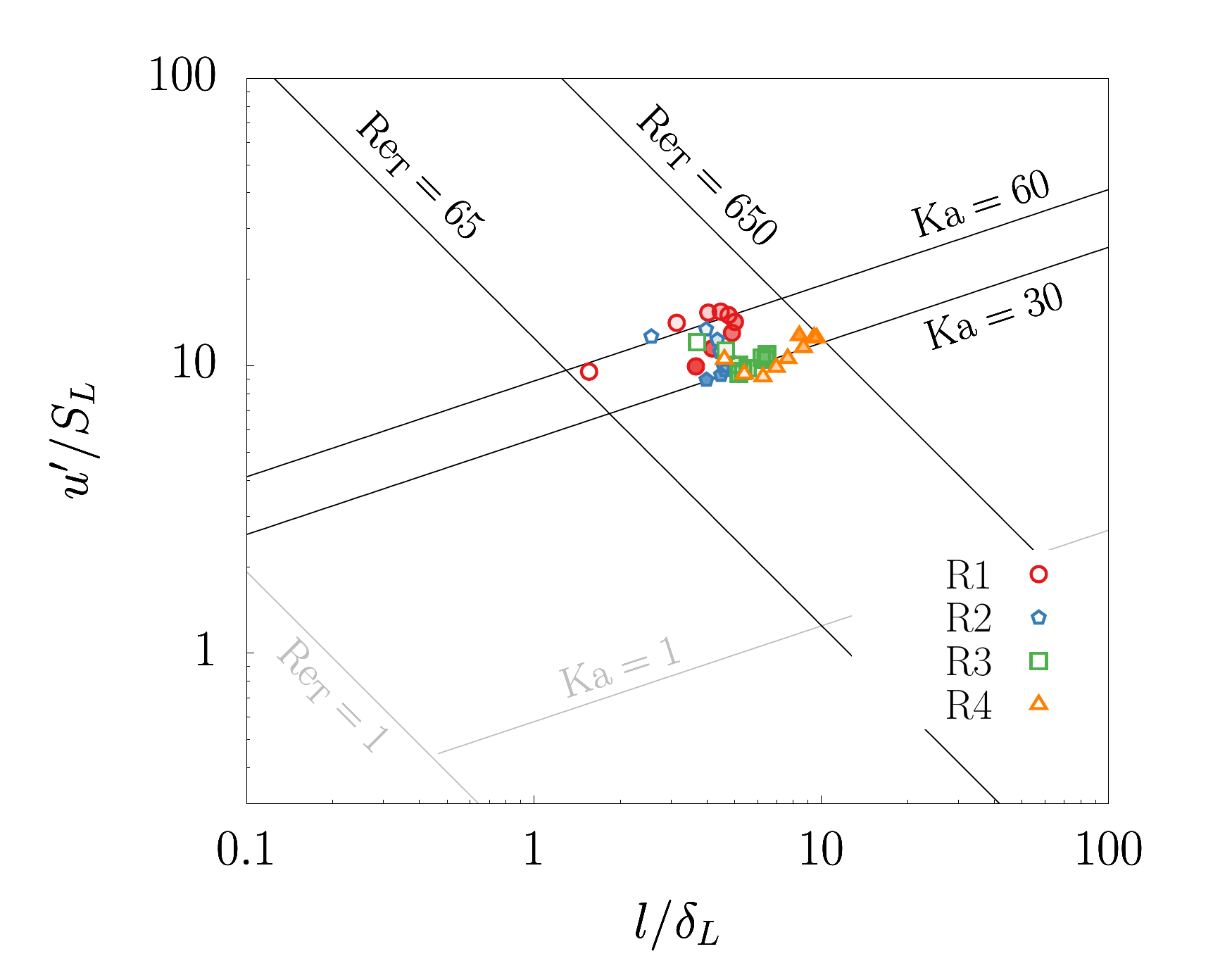}
\caption{\added{Borghi-Peters diagram for the four flames.
At each axial location, the value of the turbulent velocity fluctuations and integral scale are shown 
at the lateral location where $\ave{c}=0.73$. 
Light to dark color of the symbol filling indicates upstream to downstream positions.
}}
\label{fig:borghi}
\end{figure}

\begin{figure*}[]
\centering \includegraphics[width=130mm]{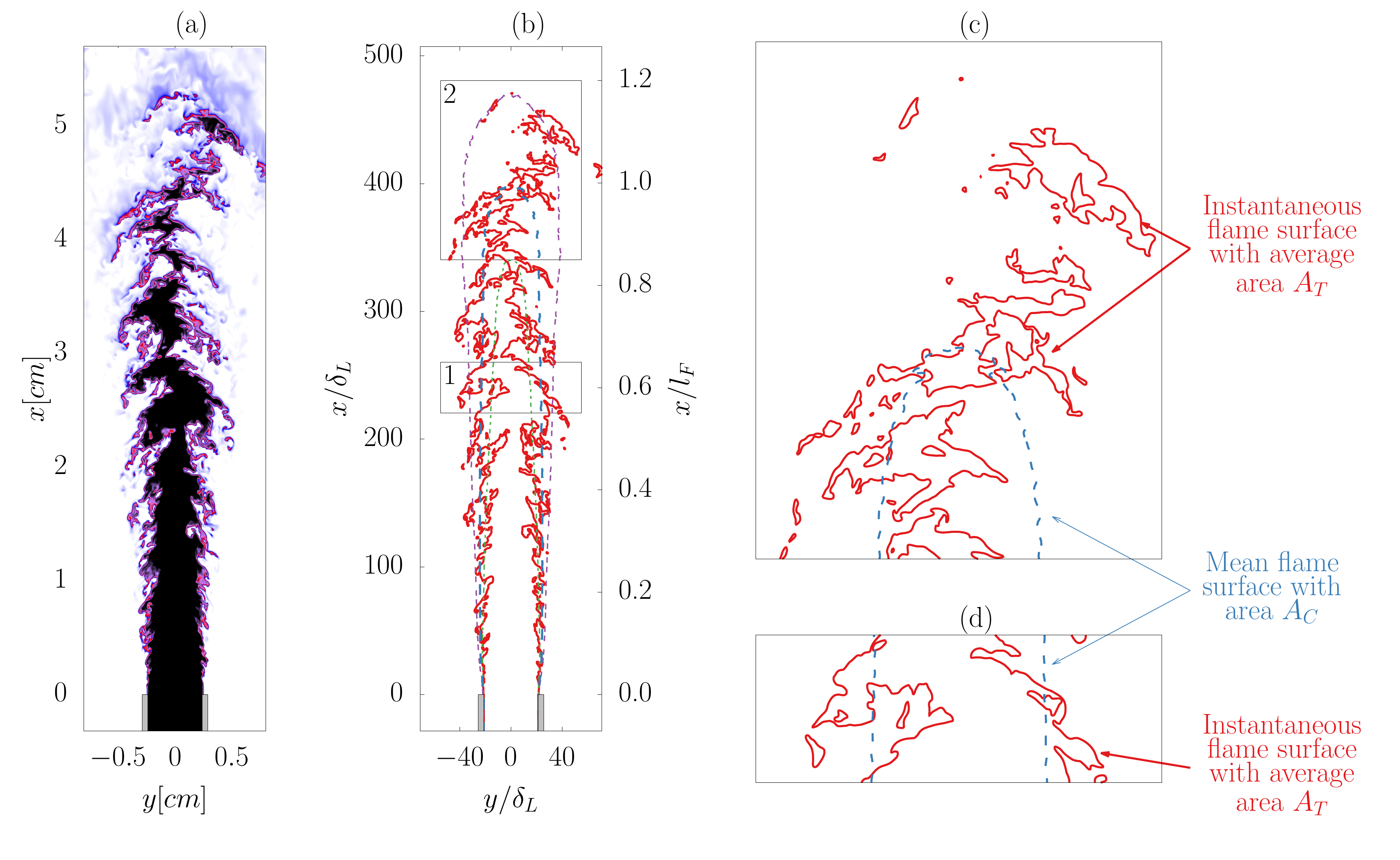}
\caption{(a) Visualization of the temperature field in a two-dimensional cut
for the flame with the highest Reynnolds number (R4). The temperatures range from 800 K
in the center of the jet to 2218 K in the coflow. 
(b) Isosurface of the instantaneous progress variable $c=0.73$ (red solid line)
and three isosurface ($\ave{c}=0.5$ green dotted line, $\ave{x}=0.73$ blue dashed line, $\ave{c}=0.9$
purple dotted line) of the mean progress variable field. 
The axes are shown in physical units in (a) and normalized with the laminar flame
thermal thickness $\delta_L$  in (b), where the streamwise direction $x$ is shown also
normalized with the flame length $l_F$.
(c,d) Enlargement of the volumes 1 and 2 in (b)
showing the instantaneous and mean flame surface. Results are shown for the
R4 flame.}
\label{fig:flame}
\end{figure*}

The flame configurations were selected in order to vary the Reynolds number, while
keeping the Karlovitz number $\Ka=\delta_L^2/\eta^2$ approximately constant.
Because the turbulence intensity is a fraction of the bulk velocity with only a minor
dependence on the Reynolds number, this scaling approach results
in a nearly constant value of $u'/S_L$ with increasing $l/\delta_L$ as the
integral scale $l$ increases with the size of the jet width $H$.
\added{
Figure~\ref{fig:borghi} shows the Borghi-Peters regime diagram for the flames. 
Following \citet{hawkes2012petascale}, the lines at constant Karlovitz $\Ka$ and turbulent Reynolds number $\Ret=u' l/\nu$ are plotted 
considering that $S_L \delta_L/\nu \approx 5.2$ for the unburned
mixture and thermodynamics conditions of the present DNS and not unity as usually assumed to draw the diagram~\citep{peters2000turbulent}.
The diagram shows that the Karlovitz number is approximately constant, especially for the flames at high Reynolds 
number and large integral scale, while the integral scale and Reynolds number increase. 
}

\begin{table}
\centering
{\footnotesize
\begin{tabular}{lcccc}
\hline
             & R1 & R2 & R3 & R4 \\
\hline 
H (mm)       & 0.6  & 1.2 & 2.4 & 4.8 \\
$\Re$        & 2800 & 5600 & 11200 & 22400 \\
\hline
$u'$  (m/s)                             & 14.3 & 10.1 & 9.9  & 11.7 \\
$l$  (mm)                               & 0.54 & 0.54 & 0.67 & 1.1  \\
$\eta$  ($\mu$m)         & 18   & 23   & 25   & 25   \\
\hline
$u'/S_L$                                & 14.2 & 10.0 & 9.8 & 11.6 \\
$l/\delta_L$                            & 4.8 & 4.8 & 5.9 & 9.5 \\
$\Ka$                                   & 39   & 23   & 21  & 21   \\
$\Rel$                                  & 49   & 39   & 40  & 50   \\
\hline
$N_x$        & 720  & 1440 & 2880 & 5760 \\
$N_y$        & 480  &  960 & 1920 & 3840 \\
$N_z$        & 256  &  256 &  512 & 1024 \\
\hline
\end{tabular}
}
\caption{Simulations parameters, evaluated at the crosswise location where $\ave{c}=c_0=0.73$
and streamwise location $x/l_F=0.6$. $\eta$ is the Kolmogorov scale, $\Rel$ the Reynolds number based
on the Taylor microscale, and $N_x$, $N_y$, $N_z$ the number of grid points in the three directions.}
\label{table}
\end{table}

The size of the computational domain is $24H$ in the streamwise ($x$),
$16H$ in the crosswise ($y$), and $4.3H$ in the spanwise ($z$) direction
($8.5H$ for R1). 
The domain is periodic in $z$, open boundary conditions are 
prescribed at the outlet in $x$ and no-slip conditions are imposed at the boundaries in $y$.
The inlet conditions for the velocity field are obtained from four auxiliary simulations of
fully developed turbulent channel flow. 

The reactive, unsteady Navier-Stokes equations are solved in the low Mach
number limit~\cite{tomboulides1997numerical}.
All transport properties are computed with a mixture-average approach~\citep{attili2016effects}
and a skeletal methane
mechanism with 16 species and 72 reactions~\cite{luca2017comprehensive} is employed.
The resolution is such that $\delta_L/\Delta \sim 6$ and $\Delta/\eta<2$ at all times.
Cases R1 and R2 were simulated also with double the spatial resolution
($\Delta = 10$~$\mu$m) without any significant change in the statistics. 
A discussion of the resolution requirements and an assessment of the quality of the solution
is discussed by \citet{luca2018statistics}.

A two-dimensional cut of the temperature field is shown in Fig.~\ref{fig:flame}a for
flame R4, characterized by the highest Reynolds number. 
An instantaneous isosurface for the value of 
the progress variable $c$ corresponding to the peak
methane mass fraction consumption rate in a one-dimensional 
unstreatched laminar flame is shown in Fig.~\ref{fig:flame}b.
The progress variable is based on the
methane mass fraction $c=1-(Y_{\ce{CH4}}/{Y}_{\ce{CH4},in})$, where 
${Y}_{\ce{CH4},in}$ is the mass fraction of methane in the unburnt mixture injected in the central jet.
 
It is evident that the flame surface is characterized by wrinkling over a large range of scales and the
overall flame size, which exceeds 400 laminar flame thicknesses in the streamwise direction, is large.
Three isocontours of the mean progress variable are also shown to identify the flame brush and its growth 
in the streamwise direction.
After a transitional region close to the jet nozzle, the flame brush grows linearly 
up to about $80\%$ of the flame length $l_F$,
defined as the position on the centerline where the mean reaction rate has the maximum. After this points, a very strong curvature of the isocontour of the mean field
is observed and the flame brush becomes about one order of magnitude larger than before. 
Along its length, the flame can be divided in three main regions: {\it (i)} a flame base region where the
Kelvin-Helmholtz instability~\citep{attili2013fluctuations} develops into turbulence, {\it (ii)}
a fully developed turbulent flame characterized by rather small mean curvature, 
{\it (iii)} a flame tip region. While the separation between these three regions is clear for the
high Reynolds number case shown in Fig.~\ref{fig:flame}, the flames at lower Reynolds numbers
show a progressive reduction in the size of the central, fully developed, turbulent region and for 
case R1, the transitional region near the base overlaps with the flame tip~\citep{luca2016direct,luca2018statistics}.

\section{Definition of the turbulent flame speed}
\label{sec:ST}

Due to the strong spatial inhomogeneity of the flow and turbulent flame
in the streamwise direction $x$, it is appropriate to define a local turbulent flame speed  
which evolves in the streamwise direction. In the present study, the turbulent flame speed is defined as 
a turbulent consumption speed, based on the reaction rate of the methane mass fraction. 
First, the streamwise direction is divided in a number of volumes 
$\mathcal{V}(x)$. Each of these extends along the entire spanwise $z$ and crosswise direction $y$.
The size of these volumes in the streamwise direction is as small as possible, yet guarantees converged statistics for the
area of isosurfaces, which is needed in the analysis.
Two examples of (the 2D projection of) these volumes are shown in Fig.~\ref{fig:flame}b by the 
boxes marked with the numbers 1 and 2.
Figures~\ref{fig:flame}c,d show enlarged views of the two regions, including a two-dimensional
cut of the instantaneous flame surface, defined as the isosurface of the methane-base progress variable 
field. The value of the progress variable $c=c_0=0.73$ is selected as it identifies the flame surface 
corresponding to the peak methane consumption in a one-dimensional unstreatched laminar flame.  
The area of the flame surface in the volume 
$\mathcal{V}$ defines an instantaneous flame area $\mathcal{A}$ and its
ensemble mean is the average flame surface area $A_{T}$.
A two-dimensional projection of a reference area $A_{C}$
is also shown in Fig.~\ref{fig:flame}(c-d); $A_{C}$ is defined from an isosurface of the mean
progress variable field using the same value used to define the instantaneous flame surface.

For volume $\mathcal{V}$, the turbulent flame speed
is defined as the integral:
\begin{equation}
S_T(x) = -\frac{\int_\mathcal{V} \rho \dot{Y}_{\ce{CH4}} dv}{\rho_u {Y}_{\ce{CH4},in} A_{C}}
\label{eq:ST}
\end{equation}
where $\rho$ and $\dot{Y}_{\ce{CH4}}$ (in units of time inverse) are the local values of the density
and reaction rate of fuel, $\rho_u$ is the unburnt gas density, ${Y}_{\ce{CH4},in}$ the
fuel mass fraction in the unburnt mixture. 

While the integral of the reaction rate is defined easily for any volume,
the reference area $A_{C}$ requires care. 
Figure~\ref{fig:flame} illustrates the way in which the reference area
is defined in the different regions of the flame. For $x/l_F \le 0.8$, the identification of an appropriate reference area for a given volume 
does not pose issues since the mean flame surface is slightly curved. 
On the contrary, at the tip, 
the mean flame surface is oriented in the crosswise direction, so that a volume of limited streamwise size would comprise only a small amount of 
the instantaneous surface.
For this reason, 
a single volume comprising the entire flame tip region (see region 2 in Fig.~\ref{fig:flame}) is considered and one
value of the turbulent flame speed is computed for $x/l_F \ge 0.8$.

The turbulent flame speed can be related to the area ratio
of the turbulent and mean flame surfaces:
\begin{equation}
\frac{S_T}{S_L} = I_0 \frac{ A_{T}}{ A_{C}}.
\end{equation}
where the coefficient $I_0$ accounts for possible deviations 
from the proportionality between the increase of flame area due to turbulent stirring 
and the turbulent flame speed.
It is worth noting that the computation of the coefficient $I_0$ 
does not require the mean reference area:
\begin{equation}
I_0 = \frac{S_T}{S_L} \frac{ A_{C}}{ A_{T}} = \frac{\Omega^*}{S_L}\frac{1}{A_{T}}
\label{eq:I_0}
\end{equation}
where
\begin{equation}
\Omega^* = -\frac{\int_\mathcal{V} \rho \dot{Y}_{\ce{CH4}} dv}{\rho_u {Y}_{\ce{CH4},in}}
\label{eq:om}
\end{equation}
is a normalized total volumetric burning rate ($m^3/s$) in the volume $\mathcal{V}$.
In addition, the issue mentioned above for an appropriate definition of the turbulent flame speed
in the flame tip region does not affect the determination of $I_0$ and the definition 
of $\Omega^*$, which may be
be computed and analyzed in volumes with a small extent in the streamwise direction in the tip region, also.


\section{Results}\label{sec:results}

\begin{figure*}[]
\centering \includegraphics[width=\two]{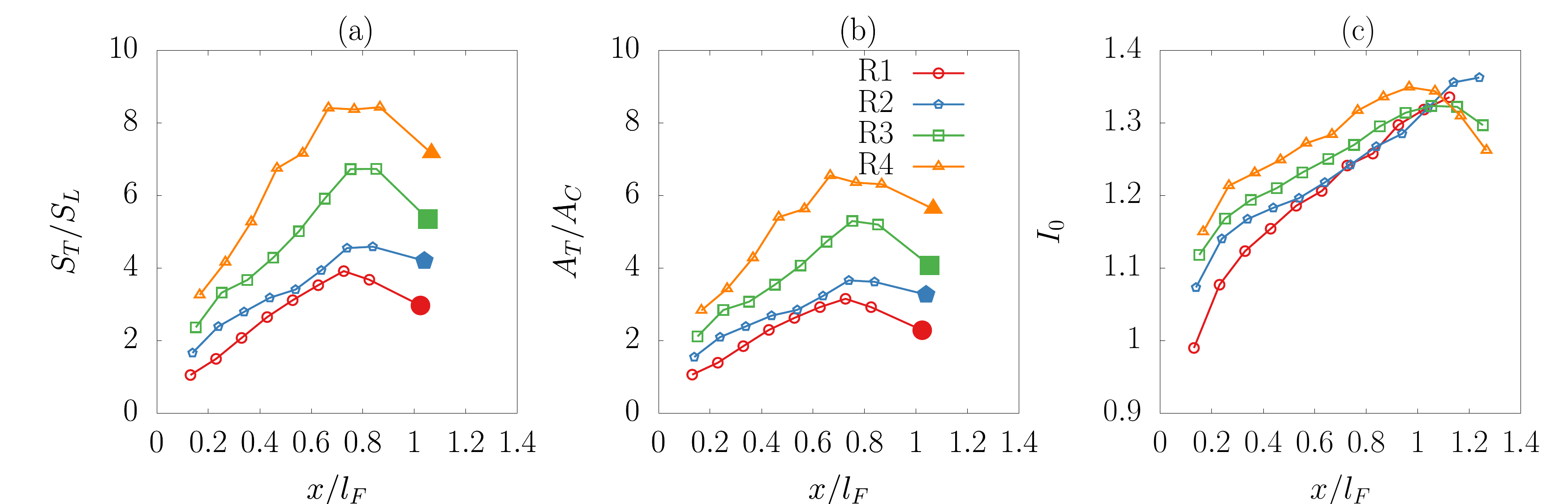}
\caption{Evolution of the turbulent flame speed (a), ratio of the turbulent and reference flame area (b), and factor 
$I_0=(S_T/S_L)/(A_T/A_C)$ (c) in the streamwise position $x$, normalized with the flame length $l_F$, for the four flames at different 
Reynolds numbers: R1 Re=2800 (red circles), 
R2 Re=5600 (blue pentagons), R3 Re=11200 (green squares), R4 Re=22440 (orange triangles). The filled symbol at $x/l_f=1$ for each flame 
corresponds to statistics computed in the large region around the flame tip identified by the box number 2 in Fig.~\ref{fig:flame}.} 
\label{fig:ST_and_I}
\end{figure*}

\begin{figure}[]
\centering \includegraphics[width=\one]{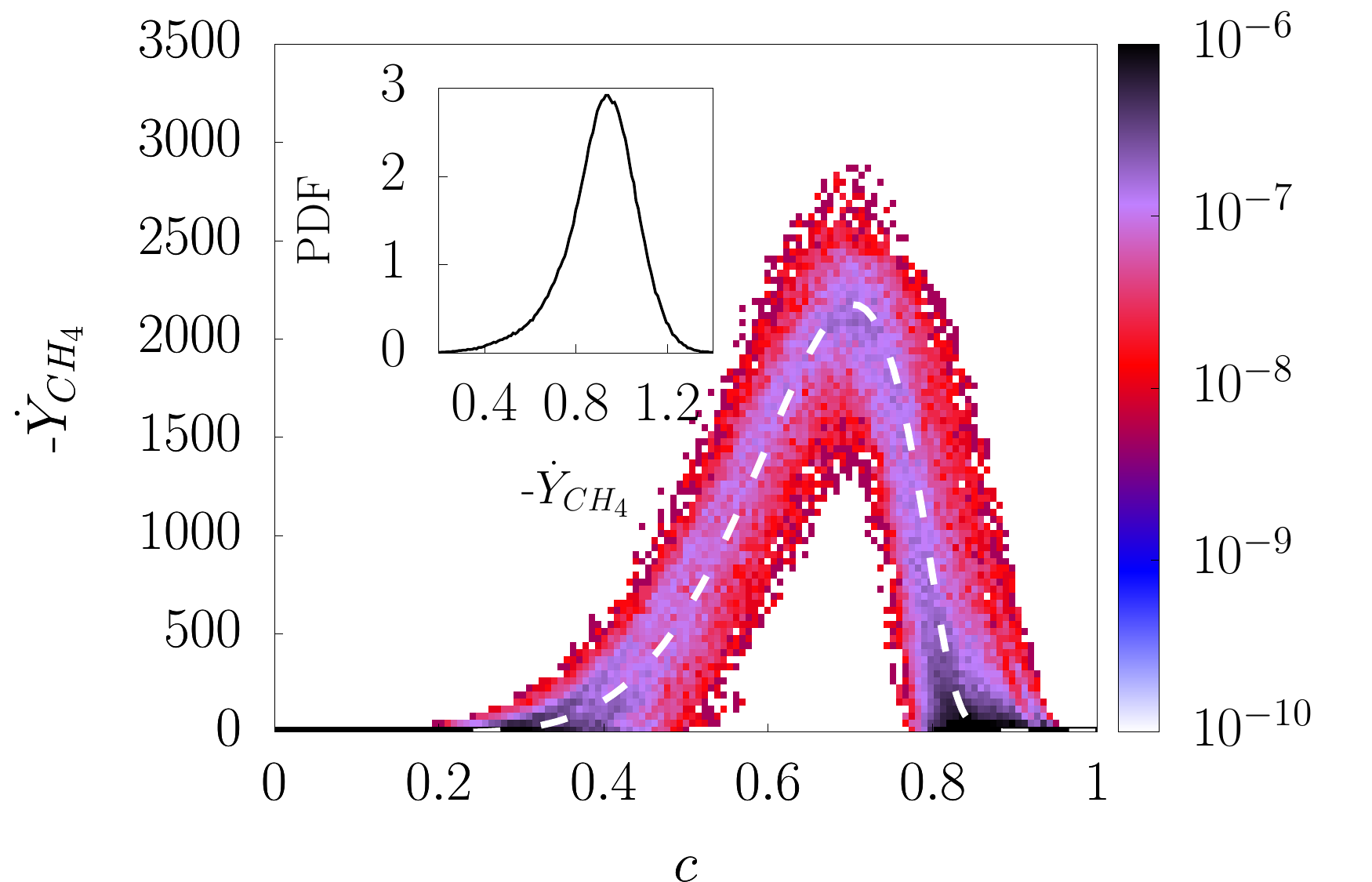}
\caption{Joint probability density function of the progress variable c and the reaction rate of methane
in flame R4 characterized by the highest Reynolds number. The white dashed line is the methane reaction 
rate in a one-dimensional laminar flame (flamelet) at the same condition of the turbulent DNS.
The inset shows the probability density function of the reaction rate conditioned on the progress
variable being $c=0.73$, corresponding to the value where the conditional mean reaction rate peaks.}
\label{fig:ReactRate_pdf}
\end{figure}

The turbulent flame speed computed with Eq.~(\ref{eq:ST}) and normalized
with the laminar flame speed $S_L$ is shown in Fig.~\ref{fig:ReactRate_pdf}a
for all four flames. The data point at $x/l_F=1$ reflects the contribution of the entire
flame tip region $x>0.8l_F$ and is marked with a different symbol accordingly. 
The flame speed increases significantly in the streamwise
direction, it reaches a maximum at $x=0.8l_F$ and then decreases slightly
to the value at the flame tip.
In addition, the flame speed increases
monotonically across flames at increasing Reynolds number.
Fig.~\ref{fig:ReactRate_pdf}b shows the 
ratio between the ensemble-averaged turbulent flame area $A_T$
and the reference area $A_C$. 
The increase of the burning velocity in the streamwise direction and with Reynolds number correlates
well with the increase in the area. Consistently, the final drop of the burning velocity at $x>0.8l_F$ is related to the
decrease of flame area. The streamwise evolution of the flame area in these flames has been explained in 
Ref.~\citep{luca2018statistics}; the increase is related to the
predominance of the strain term in the total flame stretch, 
while for $x>0.8l_F$, the propagative-curvature term overcomes strain causing the area to decrease.

It is interesting to note that, while clearly correlated, the turbulent flame speed and the area ratio
$A_T/A_C$ are not proportional. This is more evident in 
Fig.~\ref{fig:ReactRate_pdf}c where the ratio $I_0$ between the turbulent flame speed and the area is plotted.
As discussed before, the issue regarding a proper definition of the reference area $A_C$ in 
the flame tip is not relevant here because $I_0$ is defined as Eq.~\ref{eq:I_0} without computing
the reference area. For this reason, the results for $I_0$ are available up to $x=1.2l_F$ and with
high resolution in the streamwise direction also in the flame tip region.
The analysis of Fig.~\ref{fig:ReactRate_pdf}c shows that $I_0$ is always larger than one, indicating an enhancement that is not explained by
the area ratio $A_T/A_C$ alone.
In addition, the ratio $I_0$ increases in the streamwise direction and for increasing Reynolds numbers
despite the Karlovitz number remaining nearly constant.

\added{In a previous analysis of the same database performed by \citet{luca2018statistics},
it was observed that a global definition of the turbulent flame speed and 
area indicated a value for the factor $I_0$ very close to unity.
The global evaluation was based on the assumption of a statistically stationary envelope flame, which implies 
that (i) the total fuel reaction rate in the entire domain is equal 
to the flux of fuel from the inlet and (ii) the total mass flux across the flame surface is equal to the flux 
of the reactants from the inlet.  
The latter assumption might be inaccurate in high $\Ka$ flames and the global indicator might lead 
to incorrect interpretation of the physics in certain circumstances,
when mixing between the fresh gas (jet) and the fully burnt product (coflow) is present, as shown by \citet{wabel2019reaction}. 
}

From the definition of the turbulent flame speed, Eqs.~\ref{eq:ST} and~\ref{eq:om}, 
two phenomena may be responsible for the departure of $I_0$ from unity: 
{\it (i)} modifications to the reaction rate $\dot{Y}_{\ce{CH4}}$, 
{\it (ii)} thickening of the inner reaction layer of the flame.
While these two phenomena may be tightly connected because
the reaction rate tends to decrease in regions of high curvature, which 
correlate with lower gradients and flame thickening, their joint statistics are not trivial
and more analysis is required.

\begin{figure}[]
\centering \includegraphics[width=\one]{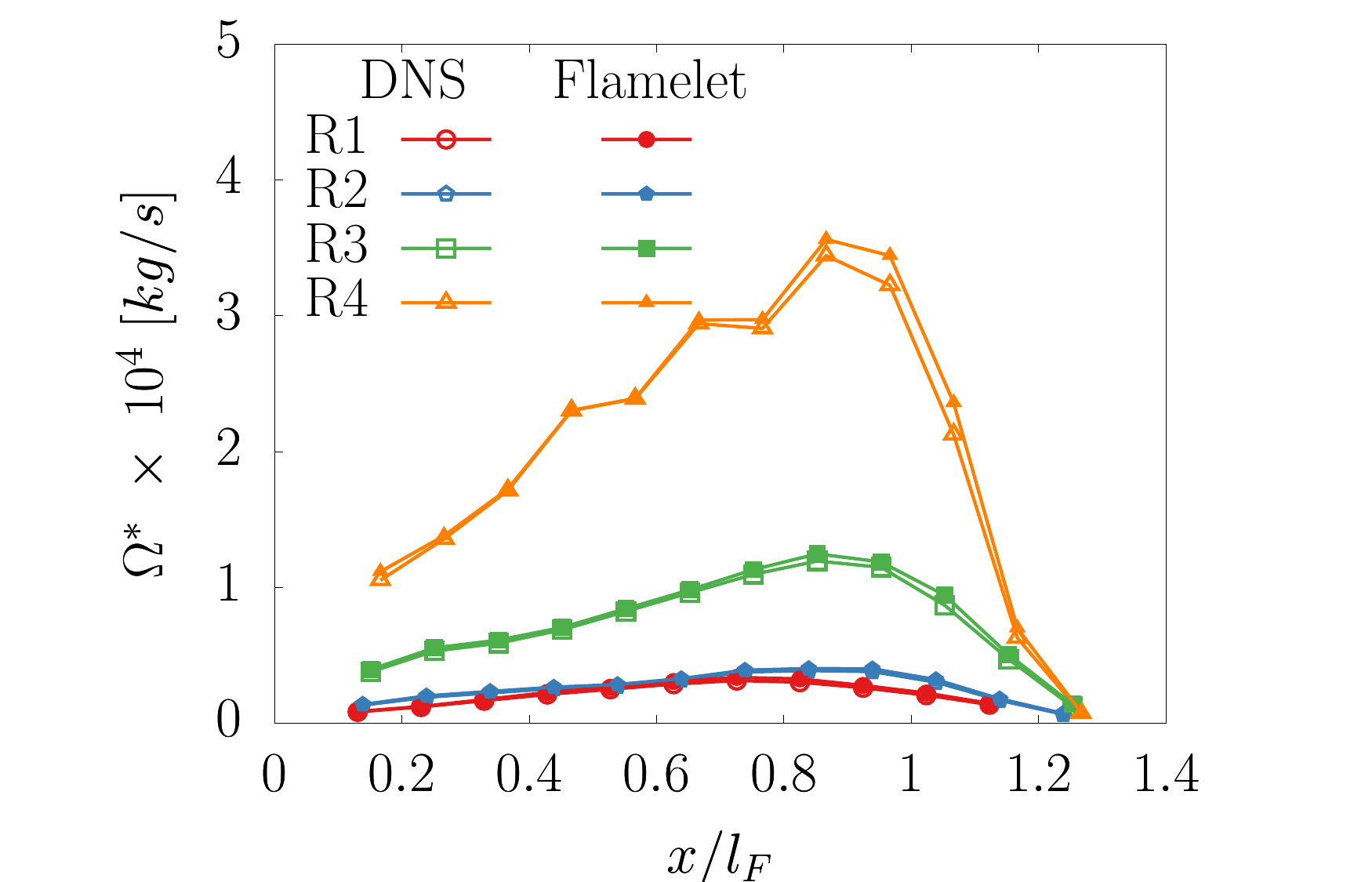}
\caption{Comparison of the integrated reaction rate $\Omega^*$ (Eq.~\ref{eq:om}) obtained from DNS data and using
a flamelet model for the local reaction rate.}
\label{fig:ST_lam}
\end{figure}

Fig.~\ref{fig:ReactRate_pdf} shows the joint probability density function (JPDF) 
of the progress variable c and methane reaction rate $\dot{Y}_{\ce{CH4}}$.
Also shown is a comparison with a flamelet obtained in a one-dimensional freely propagating
unstretched laminar flame. First, it is observed that the scatter introduced by turbulence 
is rather limited. The JPDF decreases rapidly moving away from the most probable 
value at each progress variable. This is highlighted in the inset of the same figure, 
showing the pdf of the methane reaction rate conditioned on $c=0.73$, where the peak reaction
is located. Again, the pdf is rather narrow with probability of deviation 
of $\pm 20\%$ from the most probable value
being less then $10\%$ of the maximum probability.
The comparison with the laminar flamelet highlights that the highly probable values in the DNS at each
progress variable practically matches the flamelet solution. 
This analysis suggests that the variation of $\dot{Y}_{\ce{CH4}}$
is likely to play only a minor role. This is verified in Fig.~\ref{fig:ST_lam}, where the 
quantity $\Omega^*$ (Eq.~(\ref{eq:om})), is compared with the value obtained 
computing the integral in Eq.~\ref{eq:om} with the flamelet reaction rate (white dashed line in Fig.~\ref{fig:ReactRate_pdf}) in 
place of the actual DNS value $\dot{Y}_{\ce{CH4}}$. 
The comparison shows that the
effect of turbulence on $\dot{Y}_{\ce{CH4}}$ has a negligible impact on the overall
fuel consumption rate and therefore a negligible effect on the turbulent flame speed.
The value computed with the flamelet reaction rate is slightly larger than that computed
from the actual DNS; therefore, if any, the effect of turbulence on the local reaction rate
induces $I_0$ to drop below unity, as it is usually observed in
very high Karlovitz flames~\citep{wang2017comparison,hawkes2017}.

\begin{figure}[]
\centering \includegraphics[width=\one]{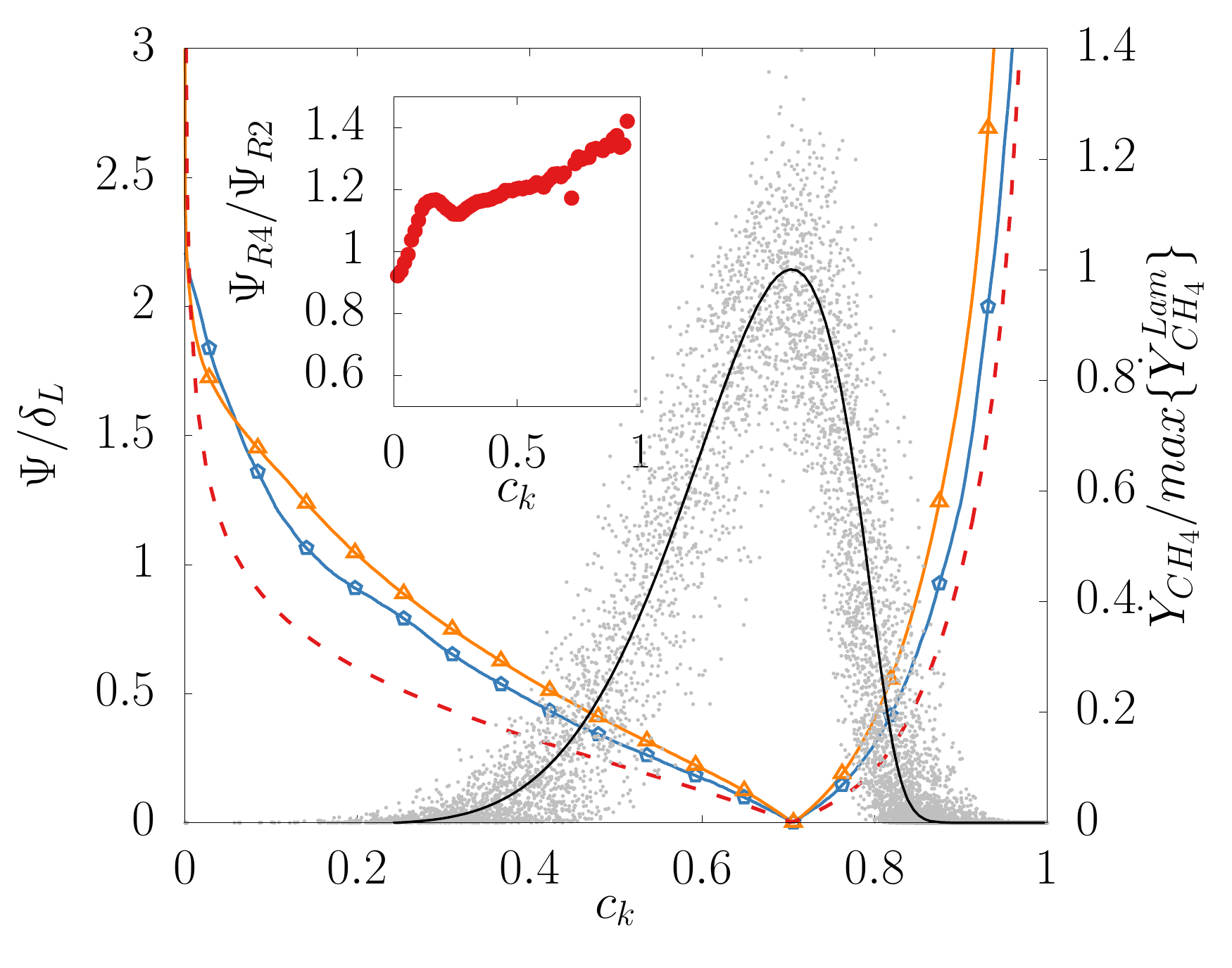}
\caption{Volume of the fluid between the flame isosurface $c=0.73$ and isosurface $c_k$ as a function of $c_k$,
normalized by the area of the isosurface $c=0.73$ and the laminar flame thickness. The results for the R2 (blue pentagons) and R4 (orange triangles) flame
are compared with those in a laminar flame (red dashed line). The scatter represents the reaction rate of methane and identifies 
the reaction layer, while the solid black line is the reaction rate in a laminar 1D flame, both normalized with the
peak in the laminar flame $max\{\dot{Y}_{CH_4}^{Lam}\}$. The
inset shows the ratio between the R4 and R2 cases. 
Data are from the volume marked as 1 in Fig~\ref{fig:flame} and spanning the streamwise locations $0.55 < x/l_F < 0.65$.
}
\label{fig:volume}
\end{figure}

From the discussion above, the value of $I_0 > 1$ must be related to a thickening of the
reaction layer of the flame that is not associated with a decrease of the reaction rate.
In order to quantify this effect, we consider a region of the flow, e.g., 
one of the volume used for the turbulent flame speed in Fig.~\ref{fig:flame}, and 
compute the volume of fluid between the flame surface,
identified by $c=c_0=0.73$, and a second isosurface $c=c_k$; then,
the volume of fluid is normalized by the area of the flame surface to obtain a
measure $\Psi$ of the distance between the flame and the isosurface $c=c_k$.
In particular,
$\Psi = 1/A_T\int_\mathcal{V}  \mathcal{H}(c_0 - c({\bf x},t)) \mathcal{H}(c({\bf x},t) - c_k) dv$ for $c_k<c_0$ and 
$\Psi = 1/A_T\int_\mathcal{V}  \mathcal{H}(c({\bf x},t) - c_0) \mathcal{H}(c_k - c({\bf x},t)) dv$ for $c_k>c_0$,
where $\mathcal{H}$ is the Heaviside function.
Values of $\Psi$ are shown in Fig.~\ref{fig:volume} for the flames R2 and R4.
The same analysis is shown also for a 1D planar laminar flame, taken to represent the baseline in the absence of thicknening induced by turbulence. 
The figure shows significant thickening for R2 and R4
in the inner layer of the flame ($0.5<c_k<0.8$), in addition to the well-documented thickening
of the preheat region observed in the range of Karlovitz numbers characteristic of the present flames.
Even more interesting is the observation that the thickening increases with Reynolds number. In particular,
the inset shows that the ratio between the thickness observed in the R4 (Re=22400) and R2 (Re=5600) flames 
is significantly above one.

\begin{figure}[]
\centering \includegraphics[width=75mm]{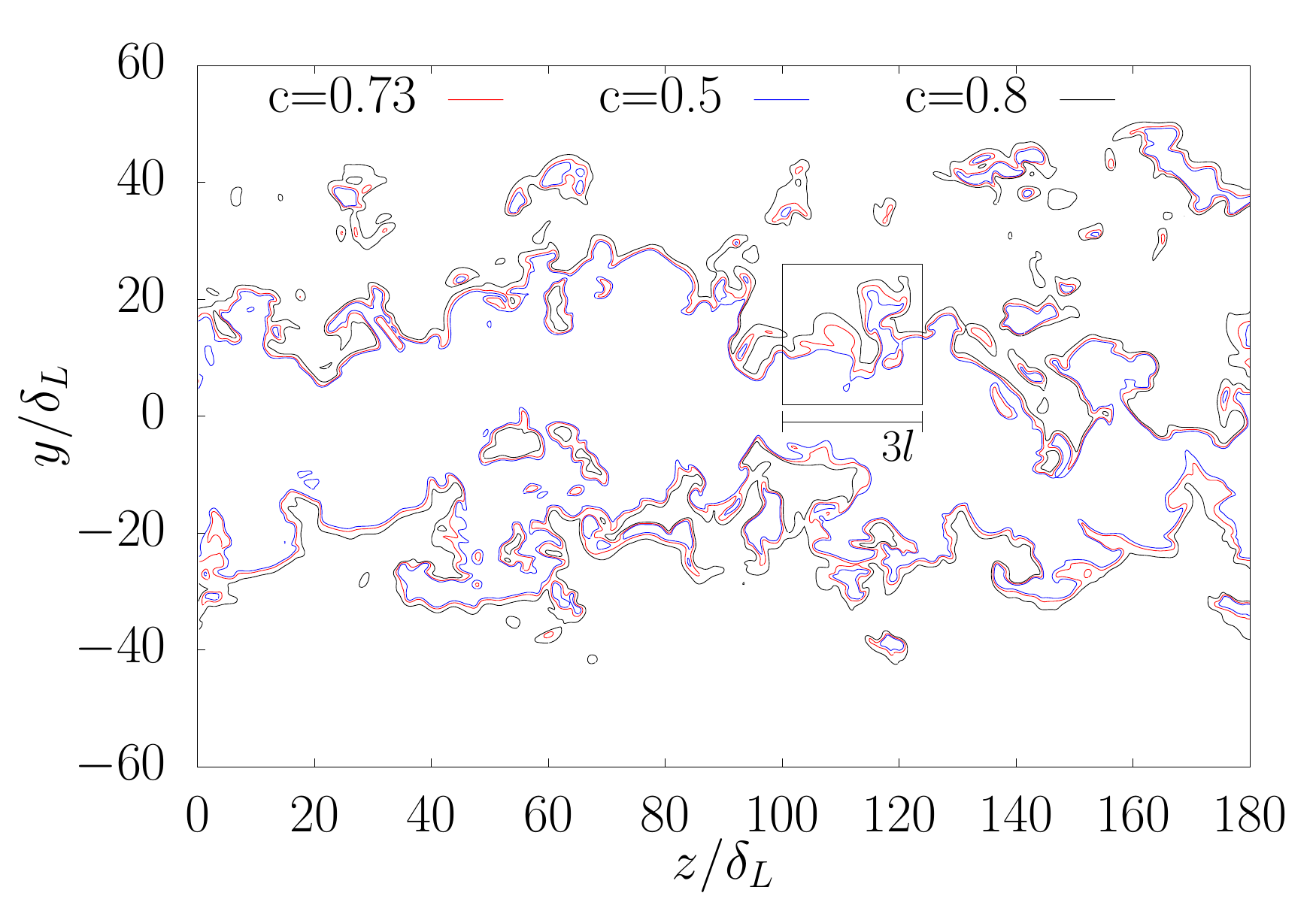}
\caption{Visualization of the progress variable field in the $y-z$ plane at $x=0.6l_F$
for the highest Reynolds number flame (R4). The isocontours 
of the three value of progress variable corresponding to the flame surface, i.e., maximum reaction rate, ($c=0.73$, red)
and to about $20\%$ of the peak reaction rate on each side of the flame  ($c=0.5$, blue and  $c=0.8$, black) are
plotted.
The box marks a region, of size $3l$, where significant thickening of the reaction layer is observed.}
\label{fig:vol_vis}
\end{figure}

The visualization of the progress variable field in Fig.~\ref{fig:vol_vis} shows a region,
marked by the black box, where a significant thickening of the reaction layer is present. 
\added{
While for clarity the visualization is shown for a two-dimensional cut, it has been
verified that the highlighted structures are not thick because of out-of-plane
effects.
}
It is worth noting that the thickened region has a size of the order of the integral scale. 
Other thickened regions are also evident and have a similarly large size. 
This observation suggests that the thickening events are related to relatively large and
intermittent turbulent structures~\citep{fiscaletti2016scale} which contain enough energy to modify the 
inner flame thickness while not affecting the reaction rate significantly.
Since the average size of the largest turbulence structure ($l$) increases with
$\Re$ at constant $\Ka$, these regions characterized by thickening get
larger and thicker for higher $\Re$, consistently with the increase of the 
normalized volume $\Psi$ shown in Fig.~\ref{fig:volume}.

\begin{figure}[]
\centering \includegraphics[width=75mm]{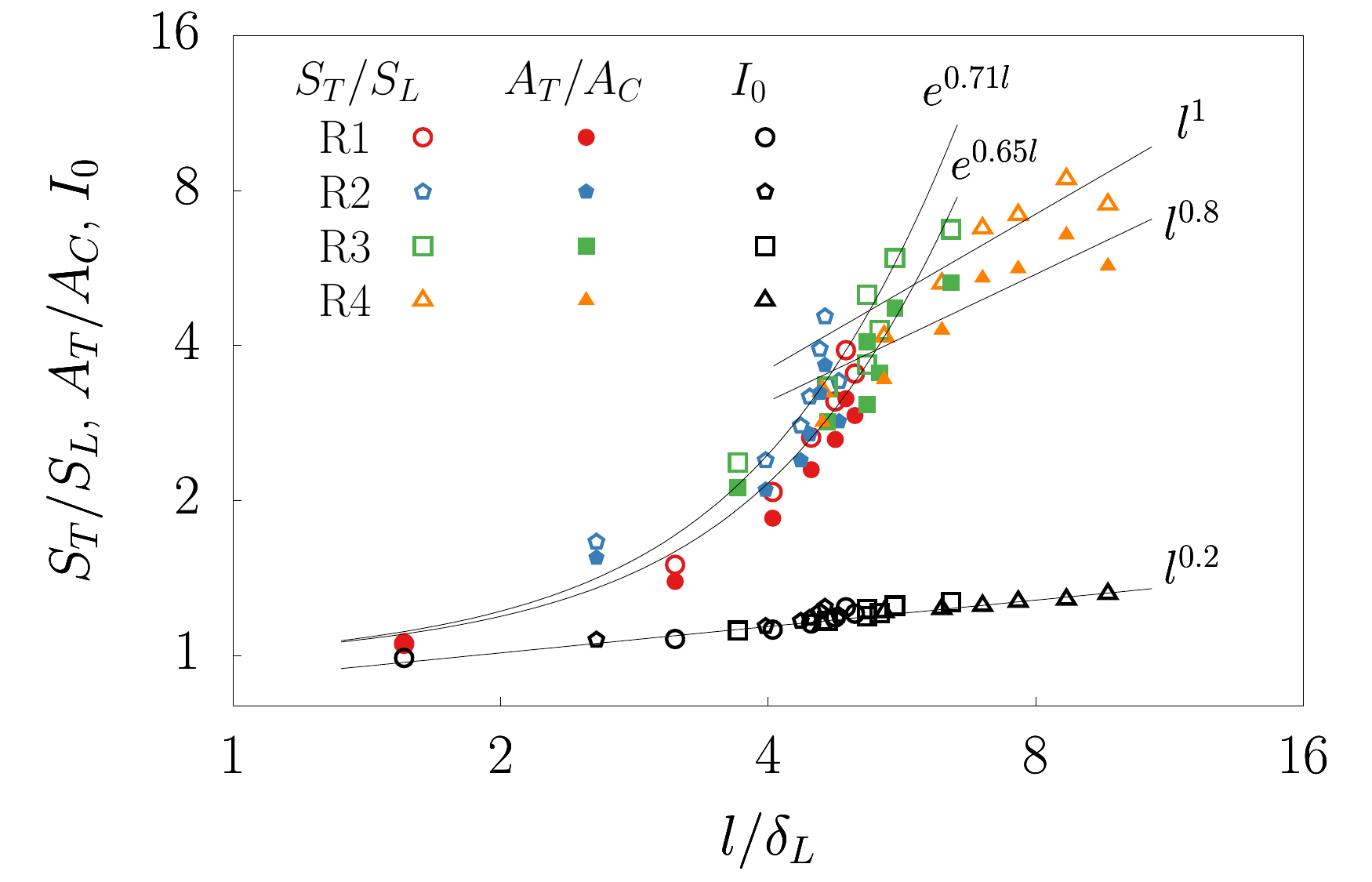}
\caption{Turbulent flame speed (color open symbols), ratio of the turbulent and reference flame area 
(color filled symbols), and factor 
$I_0=(S_T/S_L)/(A_T/A_C)$ (black open symbols) versus the integral scale of turbulence $l$ for flames at different 
Reynolds numbers: R1 Re=2800 (red circles), 
R2 Re=5600 (pentagons), R3 Re=11200 (squares), R4 Re=22440 (triangles). The lines are exponential and power-law
fits of the form indicated in the figure.}
\label{fig:ST_int_Scale}
\end{figure}

All flames in the present study
are characterized by similar turbulence intensity $u'\approx 10 S_L$ and similar Karlovitz number $\Ka=20$, so
the only parameter responsible for the Reynolds number variation across simulation is
the integral scale of turbulence $l$, which is varied by changing the jet width $H$.
A direct analysis of the effect of the integral scale is shown in Fig.~\ref{fig:ST_int_Scale}.
The turbulent flame speed $S_T/S_L$, area ratio $A_T/A_C$, and the factor $I_0$ 
are plotted versus the local integral scale for all available streamwise locations
and flames. The integral scale is defined for each streamwise location at
the crosswise position where $\ave{c}=0.73$. 
As expected, the flame speed and area ratio grow with the integral scale~\citep{peters1999turbulent,lapointe2016simulation,kulkarni2020reynolds}.
In an attempt to assess a scaling law, several curve fits for the flame speed and area ratio with respect to $l$
are included in the figure. It is observed that for relatively small integral scales, up to 
a value of $l$ corresponding to about 6 laminar flame thicknesses, the flame speed and area ratio increases
exponentially. For larger values of $l$, a power-law fitting is more appropriate with a scaling 
of $l^1$ and $l^{0.8}$ for the flame speed and area ratio, respectively.
Given the power-law scaling $l^1$ and $l^{0.8}$ for large $l$, the factor $I_0$ should scale 
as $l^{0.2}$. This scaling is also reported in the figure. 
While some uncertainty in the scaling exponents of the power-law for the flame speed, area ratio, and
$I_0$ is evident, the observation that the increase of $I_0$ with
the integral scale, or equivalently, that the inner flame layer becomes thicker for larger values 
of the integral scale, appear to be conclusive. 
\added{
The same scaling is observed also if the turbulent flame speed, ratio of the turbulent and reference flame area, and the factor 
$I_0$ are plotted versus the turbulent Reynolds number $\Ret=u' l/\nu$; the plot and a brief discussion are reported in the supplementary material.
This last observation confirms that the minor variations of Karlovitz number across the flames do not play an important role.
}

\added{
The role of the integral scale $l$ at constant Karlovitz number has been recently investigated by \citet{lapointe2016simulation}
in homogeneous isotropic turbulence; in that work,
the factor $I_0$ and inner flame thickness were found to be independent of $l$.
There are several circumstances that might explain the difference. 
First, the presence of shear in the present jet configuration might play a role.
However, shear decreases with the Reynolds number and downstream distance in the present configuration and, usually,
it tends to have a decreasing impact on the dynamics of turbulence 
as the Reynolds number increases~\citep{casciola2005scaling,celani2005shear,attili2012statistics,attili2013fluctuations},
likely implying that shear might not be the reason of the observed behavior. 
Moreover, the flames considered by \citet{lapointe2016simulation} are characterized by a rather 
small ratio between the inner flame layer and the thermal thicknesses; in the present work, due to the high temperature
of the fresh mixture, the reaction layer is rather thick and therefore more prone to be affected by turbulence. 
It is then possible that the values of integral scale considered by Lapointe, ranging between 1 and 4 flame thicknesses, are too small 
to highlight the effects on $I_0$.
Finally, Fig~\ref{fig:vol_vis} shows that the regions where the thickening of the inner layer is observed are rather large,
of the order of 3 integral scales. Likely, this fact implies that a large domain is required to avoid boundary effects. 
Therefore, the limited domain size employed in the work of \citet{lapointe2016simulation}, which is 
fixed at 5 integral scales due to the linear forcing employed, might limit the effect of the integral scale on $I_0$.
Nevertheless, the data currently available do not allow a systematic assessment of all the possible
reasons of the difference and additional analyses will certainly be of interest.
}

The conclusion that the factor $I_0$ and the thickening of the inner layer increase with 
Reynolds number at nearly constant Karlovitz is a significant result. It points to the possibility that,
contrary to the classical theory describing the effects of the two main parameters of the Borghi-Peters diagram
($l$ and $u'$ or $\Re$ and $\Ka$),
an increase of $l$ or $Re$ at constant $u'$ or $\Ka$ might cause 
variations in $I_0$ and thickening of the inner layer, which are usually ascribed solely to variations in $u'$ and $\Ka$.
This result is, to a certain extent, consistent with the recent work of \citet{driscoll2020premixed} who reported
a new {\it measured} diagram of combustion in which a regime characterized
by a broadened preheat layer appears increasing the integral scale $l$ at constant turbulent intensity $u'$.

Based on the data and discussions above, we speculate that the transition from 
the thin reaction zone regime to distributed burning\citep{aspden2019towards}, in which turbulence alters transport
in the inner layer, might occur at lower values of $\Ka$ for high $\Re$ (large integral scales)
compared to the value of $\Ka$ needed to observe distributed burning at low $\Re$ (small integral scale).
In other words, the transition may depend on both $\Ka$ and $\Re$.
Direct verification of this conjecture would require additional simulations at different 
values of $\Ka$ for each $\Re$ considered in the present flame set.

\section{Conclusions}\label{sec:conclusions}
The data from four DNS of turbulent premixed slot jet flames up to $\Re=22000$
show that the turbulent flame speed, evaluated at each streamwise position, increases 
with the distance from the nozzle and with the Reynolds number of the flame. 
This is not entirely due to the increase of the flame area since
a significant thickening of the inner reaction layer,
associated with negligible changes of the reaction rates, also contributes
to the enhancement of the turbulent flame speed.
It is concluded that, even if the flame are characterized by a constant Karlovitz number,
an higher Reynolds number, and correspondingly larger turbulence integral scale,
enhances the effect of turbulence inside the inner flame layer.

\section*{Acknowledgements}
The authors acknowledge funding from of the European Research Council (ERC) under the European 
Union’s Horizon 2020 research and innovation program under grant agreement No 695747.


\bibliography{biblio.bib} 

\begin{thebibliography}{23}
\expandafter\ifx\csname natexlab\endcsname\relax\def\natexlab#1{#1}\fi
\providecommand{\bibinfo}[2]{#2}

\bibitem[{Peters(2000)}]{peters2000turbulent}
\bibinfo{author}{N.~Peters}, \bibinfo{title}{Turbulent combustion}, Cambridge
  university press, \bibinfo{year}{2000}.

\bibitem[{Veynante and Vervisch(2002)}]{veynante2002turbulent}
\bibinfo{author}{D.~Veynante}, \bibinfo{author}{L.~Vervisch},
  \bibinfo{title}{Turbulent combustion modeling}, \bibinfo{journal}{Prog.
  Energ. Combust. Sci.} \bibinfo{volume}{28} (\bibinfo{year}{2002})
  \bibinfo{pages}{193--266}.

\bibitem[{Driscoll(2008)}]{driscoll2008turbulent}
\bibinfo{author}{J.~F. Driscoll}, \bibinfo{title}{Turbulent premixed
  combustion: Flamelet structure and its effect on turbulent burning
  velocities}, \bibinfo{journal}{Prog. Energ. Combust. Sci.}
  \bibinfo{volume}{34} (\bibinfo{year}{2008}) \bibinfo{pages}{91--134}.

\bibitem[{Driscoll et~al.(2020)Driscoll, Chen, Skiba, Carter, Hawkes, and
  Wang}]{driscoll2020premixed}
\bibinfo{author}{J.~F. Driscoll}, \bibinfo{author}{J.~H. Chen},
  \bibinfo{author}{A.~W. Skiba}, \bibinfo{author}{C.~D. Carter},
  \bibinfo{author}{E.~R. Hawkes}, \bibinfo{author}{H.~Wang},
  \bibinfo{title}{Premixed flames subjected to extreme turbulence: Some
  questions and recent answers}, \bibinfo{journal}{Prog. Energ. Combust. Sci.}
  \bibinfo{volume}{76} (\bibinfo{year}{2020}) \bibinfo{pages}{100802}.

\bibitem[{Luca et~al.(2019)Luca, Attili, Lo~Schiavo, Creta, and
  Bisetti}]{luca2018statistics}
\bibinfo{author}{S.~Luca}, \bibinfo{author}{A.~Attili},
  \bibinfo{author}{E.~Lo~Schiavo}, \bibinfo{author}{F.~Creta},
  \bibinfo{author}{F.~Bisetti}, \bibinfo{title}{On the statistics of flame
  stretch in turbulent premixed jet flames in the thin reaction zone regime at
  varying Reynolds number}, \bibinfo{journal}{Proc. Combust. Inst.}
  \bibinfo{volume}{37} (\bibinfo{year}{2019}) \bibinfo{pages}{2451--2459}.

\bibitem[{Peters(1999)}]{peters1999turbulent}
\bibinfo{author}{N.~Peters}, \bibinfo{title}{The turbulent burning velocity for
  large-scale and small-scale turbulence}, \bibinfo{journal}{J. Fluid Mech.}
  \bibinfo{volume}{384} (\bibinfo{year}{1999}) \bibinfo{pages}{107--132}.

\bibitem[{Lapointe(2016)}]{lapointe2016simulation}
\bibinfo{author}{S.~Lapointe}, \bibinfo{title}{Simulation of premixed
  hydrocarbon flames at high turbulence intensities}, Ph.D. thesis, California
  Institute of Technology, \bibinfo{year}{2016}.

\bibitem[{Kulkarni et~al.(view)Kulkarni, Attili, and
  Bisetti}]{kulkarni2020reynolds}
\bibinfo{author}{T.~Kulkarni}, \bibinfo{author}{A.~Attili},
  \bibinfo{author}{F.~Bisetti}, \bibinfo{title}{Reynolds number scaling of
  burning rates in spherical turbulent premixed flames}, \bibinfo{journal}{J.
  Fluid Mech.}  (\bibinfo{year}{under review}).

\bibitem[{Skiba et~al.(2018)Skiba, Wabel, Carter, Hammack, Temme, and
  Driscoll}]{skiba2018premixed}
\bibinfo{author}{A.~W. Skiba}, \bibinfo{author}{T.~M. Wabel},
  \bibinfo{author}{C.~D. Carter}, \bibinfo{author}{S.~D. Hammack},
  \bibinfo{author}{J.~E. Temme}, \bibinfo{author}{J.~F. Driscoll},
  \bibinfo{title}{Premixed flames subjected to extreme levels of turbulence
  part I: Flame structure and a new measured regime diagram},
  \bibinfo{journal}{Combust. Flame} \bibinfo{volume}{189}
  (\bibinfo{year}{2018}) \bibinfo{pages}{407--432}.

\bibitem[{Hawkes et~al.(2012)Hawkes, Chatakonda, Kolla, Kerstein, and
  Chen}]{hawkes2012petascale}
\bibinfo{author}{E.~R. Hawkes}, \bibinfo{author}{O.~Chatakonda},
  \bibinfo{author}{H.~Kolla}, \bibinfo{author}{A.~R. Kerstein},
  \bibinfo{author}{J.~H. Chen}, \bibinfo{title}{A petascale direct numerical
  simulation study of the modelling of flame wrinkling for large-eddy
  simulations in intense turbulence}, \bibinfo{journal}{Combust. Flame}
  \bibinfo{volume}{159} (\bibinfo{year}{2012}) \bibinfo{pages}{2690--2703}.

\bibitem[{Tomboulides et~al.(1997)Tomboulides, Lee, and
  Orszag}]{tomboulides1997numerical}
\bibinfo{author}{A.~Tomboulides}, \bibinfo{author}{J.~Lee},
  \bibinfo{author}{S.~Orszag}, \bibinfo{title}{Numerical simulation of low Mach
  number reactive flows}, \bibinfo{journal}{J. Sci. Comput.}
  \bibinfo{volume}{12} (\bibinfo{year}{1997}) \bibinfo{pages}{139--167}.

\bibitem[{Attili et~al.(2016)Attili, Bisetti, Mueller, and
  Pitsch}]{attili2016effects}
\bibinfo{author}{A.~Attili}, \bibinfo{author}{F.~Bisetti},
  \bibinfo{author}{M.~E. Mueller}, \bibinfo{author}{H.~Pitsch},
  \bibinfo{title}{Effects of non-unity Lewis number of gas-phase species in
  turbulent nonpremixed sooting flames}, \bibinfo{journal}{Combust. Flame}
  \bibinfo{volume}{166} (\bibinfo{year}{2016}) \bibinfo{pages}{192--202}.

\bibitem[{Luca et~al.(2017)Luca, Al-Khateeb, Attili, and
  Bisetti}]{luca2017comprehensive}
\bibinfo{author}{S.~Luca}, \bibinfo{author}{A.~N. Al-Khateeb},
  \bibinfo{author}{A.~Attili}, \bibinfo{author}{F.~Bisetti},
  \bibinfo{title}{Comprehensive Validation of Skeletal Mechanism for Turbulent
  Premixed Methane--Air Flame Simulations}, \bibinfo{journal}{J. Propul. Power}
   (\bibinfo{year}{2017}) \bibinfo{pages}{1--8}.

\bibitem[{Attili and Bisetti(2013)}]{attili2013fluctuations}
\bibinfo{author}{A.~Attili}, \bibinfo{author}{F.~Bisetti},
  \bibinfo{title}{Fluctuations of a passive scalar in a turbulent mixing
  layer}, \bibinfo{journal}{Phys. Rev. E} \bibinfo{volume}{88}
  (\bibinfo{year}{2013}) \bibinfo{pages}{033013}.

\bibitem[{Luca et~al.(0189)Luca, Attili, and Bisetti}]{luca2016direct}
\bibinfo{author}{S.~Luca}, \bibinfo{author}{A.~Attili},
  \bibinfo{author}{F.~Bisetti}, \bibinfo{title}{Direct Numerical Simulation of
  Turbulent Lean Methane-Air Bunsen Flames with Mixture Inhomogeneities},
  \bibinfo{journal}{54th AIAA Aerospace Sciences Meeting, AIAA SciTech Forum}
  (\bibinfo{year}{AIAA 2016-0189}).

\bibitem[{Wabel et~al.(2019)Wabel, Barlow, and Steinberg}]{wabel2019reaction}
\bibinfo{author}{T.~M. Wabel}, \bibinfo{author}{R.~S. Barlow},
  \bibinfo{author}{A.~M. Steinberg}, \bibinfo{title}{Reaction zone
  stratification in piloted highly-turbulent fuel-lean premixed jets},
  \bibinfo{journal}{Combust. Flame} \bibinfo{volume}{208}
  (\bibinfo{year}{2019}) \bibinfo{pages}{327--329}.

\bibitem[{Wang et~al.(2017{\natexlab{a}})Wang, Hawkes, Zhou, Chen, Li, and
  Ald{\'e}n}]{wang2017comparison}
\bibinfo{author}{H.~Wang}, \bibinfo{author}{E.~R. Hawkes},
  \bibinfo{author}{B.~Zhou}, \bibinfo{author}{J.~H. Chen},
  \bibinfo{author}{Z.~Li}, \bibinfo{author}{M.~Ald{\'e}n}, \bibinfo{title}{A
  comparison between direct numerical simulation and experiment of the
  turbulent burning velocity-related statistics in a turbulent methane-air
  premixed jet flame at high Karlovitz number}, \bibinfo{journal}{Proc.
  Combust. Inst.} \bibinfo{volume}{36} (\bibinfo{year}{2017}{\natexlab{a}})
  \bibinfo{pages}{2045--2053}.

\bibitem[{Wang et~al.(2017{\natexlab{b}})Wang, Hawkes, Chen, Zhou, Li, and
  Ald{\'e}n}]{hawkes2017}
\bibinfo{author}{H.~Wang}, \bibinfo{author}{E.~R. Hawkes},
  \bibinfo{author}{J.~H. Chen}, \bibinfo{author}{B.~Zhou},
  \bibinfo{author}{Z.~Li}, \bibinfo{author}{M.~Ald{\'e}n},
  \bibinfo{title}{Direct numerical simulations of a high Karlovitz number
  laboratory premixed jet flame \--- an analysis of flame stretch and flame
  thickening}, \bibinfo{journal}{J. Fluid Mech.}
  (\bibinfo{year}{2017}{\natexlab{b}}) \bibinfo{pages}{511--536}.

\bibitem[{Fiscaletti et~al.(2016)Fiscaletti, Attili, Bisetti, and
  Elsinga}]{fiscaletti2016scale}
\bibinfo{author}{D.~Fiscaletti}, \bibinfo{author}{A.~Attili},
  \bibinfo{author}{F.~Bisetti}, \bibinfo{author}{G.~E. Elsinga},
  \bibinfo{title}{Scale interactions in a mixing layer -- the role of the
  large-scale gradients}, \bibinfo{journal}{J. Fluid Mech.}
  \bibinfo{volume}{791} (\bibinfo{year}{2016}) \bibinfo{pages}{154--173}.

\bibitem[{Casciola et~al.(2005)Casciola, Gualtieri, Jacob, and
  Piva}]{casciola2005scaling}
\bibinfo{author}{C.~M. Casciola}, \bibinfo{author}{P.~Gualtieri},
  \bibinfo{author}{B.~Jacob}, \bibinfo{author}{R.~Piva},
  \bibinfo{title}{Scaling properties in the production range of shear dominated
  flows}, \bibinfo{journal}{Phys. Rev. Lett.} \bibinfo{volume}{95}
  (\bibinfo{year}{2005}) \bibinfo{pages}{024503}.

\bibitem[{Celani et~al.(2005)Celani, Cencini, Vergassola, Villermaux, and
  Vincenzi}]{celani2005shear}
\bibinfo{author}{A.~Celani}, \bibinfo{author}{M.~Cencini},
  \bibinfo{author}{M.~Vergassola}, \bibinfo{author}{E.~Villermaux},
  \bibinfo{author}{D.~Vincenzi}, \bibinfo{title}{Shear effects on passive
  scalar spectra}, \bibinfo{journal}{J. Fluid Mech.} \bibinfo{volume}{523}
  (\bibinfo{year}{2005}) \bibinfo{pages}{99--108}.

\bibitem[{Attili and Bisetti(2012)}]{attili2012statistics}
\bibinfo{author}{A.~Attili}, \bibinfo{author}{F.~Bisetti},
  \bibinfo{title}{{Statistics and scaling of turbulence in a spatially
  developing mixing layer at Re$_\lambda$= 250}}, \bibinfo{journal}{Phys.
  Fluids} \bibinfo{volume}{24} (\bibinfo{year}{2012}) \bibinfo{pages}{035109}.

\bibitem[{Aspden et~al.(2019)Aspden, Day, and Bell}]{aspden2019towards}
\bibinfo{author}{A.~J. Aspden}, \bibinfo{author}{M.~S. Day},
  \bibinfo{author}{J.~B. Bell}, \bibinfo{title}{Towards the distributed burning
  regime in turbulent premixed flames}, \bibinfo{journal}{J. Fluid Mech.}
  \bibinfo{volume}{871} (\bibinfo{year}{2019}) \bibinfo{pages}{1--21}.

\end{thebibliography}
\bibliographystyle{elsarticle-num-CNF-titles.bst}


\end{document}